\documentclass[letterpaper,longbibliography,twocolumn,showpacs,floatfix,groupaddress,twocolumns]{revtex4-1}
\usepackage[latin1]{inputenc}
\usepackage{bm}
\usepackage{multirow}
\usepackage{amssymb}
\usepackage{amsbsy}
\usepackage{amsmath,  amsthm, amsfonts, mathrsfs}
\usepackage{graphicx}
\usepackage{epsfig}
\usepackage{placeins}
\usepackage{wrapfig}
\usepackage{float}
\usepackage{trfsigns}
\usepackage[usenames,dvipsnames]{color}
\usepackage[colorlinks,linkcolor=blue,citecolor=blue,urlcolor=blue,breaklinks=true]{hyperref}
\newcommand{\mcl}{\mathcal{L}}
\newcommand{\mco}{\mathcal{O}}
\newcommand{\mch}{\mathcal{H}}
\newcommand{\mcq}{\mathcal{Q}}
\newcommand{\mcv}{\mathcal{V}}
\newcommand{\mce}{\mathcal{E}}

\newcommand{\lc}{Lanczos coefficients }
\newcommand{\ii}{\mathrm{i}}

\makeatletter

\begin{document} 

 \title{Stability of Exponentially Damped Oscillations under Perturbations of the Mori-Chain}
 
 \author{Robin Heveling}
 \email{rheveling@uos.de}
 \affiliation{Department of Physics, University of Osnabr\"uck, D-49076 Osnabr\"uck, Germany}

 \author{Jiaozi Wang}
 \email{jiaowang@uos.de}
 \affiliation{Department of Physics, University of Osnabr\"uck, D-49076 Osnabr\"uck, Germany}
 
  \author{Christian Bartsch}
 \email{cbartsch@uos.de}
 \affiliation{Department of Physics, University of Osnabr\"uck, D-49076 Osnabr\"uck, Germany}

 \author{Jochen Gemmer}
 \email{jgemmer@uos.de}
 \affiliation{Department of Physics, University of Osnabr\"uck, D-49076 Osnabr\"uck, Germany}

\begin{abstract}
There is an abundance of evidence that some relaxation dynamics, e.g., exponential decays,
are much more common in nature than others. Recently, there have been attempts to trace
this dominance back to a certain stability of the prevalent dynamics versus generic Hamiltonian perturbations.
In the paper at hand, we tackle this stability issue from yet another angle, namely in the framework of the recursion method.
We investigate the behavior of various relaxation dynamics with respect to alterations of the so-called Lanczos coefficients. All considered scenarios are set up in order to comply with the ``universal operator growth hypothesis''. Our numerical experiments suggest the existence of stability in a larger  class of relaxation dynamics consisting  of exponentially damped oscillations. 
Further, we propose a criterion to identify ``pathological'' perturbations that lead to uncommon dynamics.
\end{abstract}
\maketitle

\section{Introduction}
\label{intro}
\noindent
The apparent emergence of irreversibility from the underlying reversible theory of quantum mechanics is a long-standing puzzle that lacks an entirely satisfying answer to this day \cite{gogolin}. Over the course of the last decades, fundamental concepts like the ``eigenstate thermalization hypothesis'' \cite{sred,deutsch,ales} and ``quantum typicality'' \cite{lloyd,gold,rei} have crystallized, which give conditions under which isolated quantum systems eventually reach an equilibrium state. However, while these mechanisms ensure eventual equilibration, they make no statement in which manner the equilibrium state is actually reached, i.e., they do not narrow down the eligible routes to equilibrium. In contrast, it is evidently true that some relaxation dynamics, e.g., exponential decays, are much more common in nature than others, e.g., recurrence dynamics.\\ 
There are recent attempts to explain this prevalence with the idea that some dynamics are stable versus perturbations of the Hamiltonian. These attempts include, for example, investigations based on Hamiltonian perturbations on the level of random matrices \cite{knip}. Further, there exist advances that consider an entire ensemble of permissible perturbations \cite{dabe1,dabe2,dabe3}. It is then analytically shown that weak ``typical'' perturbations lead to an exponential damping of the original dynamics \cite{dabe1}. This renders exponential decays stable since only the decay constant changes, whereas recurrence dynamics are exponentially suppressed.\\
With the paper at hand, we address the issue of stability of certain [classes of] relaxation dynamics in the framework of the \textit{recursion method} \cite{recu,recu2}. Central quantities that appear within this framework are the so-called Lanczos coefficients, real numbers that characterize the
complexity growth of operators over the course of time. As will be presented in Sec.\ \ref{set}, the Lanczos coefficients can be interpreted as hopping amplitudes in a tight-binding model. In this manner,  many physical problems, like calculating correlation functions, can practically be reduced to a one-dimensional [finite or semi-infinite] chain, which we refer to as the ``Mori-chain'' in the title of this paper. In the following, we consider perturbations on the level of Lanczos coefficients and examine the resulting effect on various kinds of relaxation dynamics.\\
The particular advantage of our approach within the recursion method framework is that all considered scenarios can be set up to directly comply with the universal operator growth hypothesis \cite{berkley}, which previous approaches have been lacking \cite{knip}. Said hypothesis concerns the asymptotic growth of the Lanczos coefficients and it basically states that the coefficients should eventually attain linear growth  [with a logarithmic correction in one dimension]. The hypothesis is backed up by analytical as well as numerical evidence \cite{berkley,dong,hev}.\\
The paper at hand is structured as follows:\ Sec.\ \ref{set} constitutes a preliminary section on the recursion method and the operator growth hypothesis. Afterwards, in \mbox{Sec.\ \ref{stabil}}, the concrete strategy to study the stability of certain classes of dynamics is explained in detail. In Sec.\ \ref{numeric} we present and discuss our numerical results. We conclude in Sec.\ \ref{conc}.

\section{Preliminaries:\ Recursion method and operator growth hypothesis}
\label{set}
\noindent
In this section, we briefly recall the basics of the recursion method \cite{recu,recu2} as well as the universal operator growth hypothesis \cite{berkley}. We consider a system described by some Hamiltonian $\mch$. An observable of interest represented by a Hermitian operator $\mco$ gives rise to a corresponding autocorrelation function
\begin{equation}
\label{cor}
C(t) = \text{Tr}[\mco(t)\mco]\,,
\end{equation}
where $\mco(t)=e^{\mathrm{i}\mch t}\mco e^{-\mathrm{i}\mch t}$ is the time-dependent operator in the Heisenberg picture ($\hbar=1$). In the following, it is convenient to work directly in Liouville space, i.e., the Hilbert space of operators, and denote its elements $\mco$ as states $|\mco)$. Elements of the Liouville space evolve under the Liovillian  $\mcl= [\mch,\cdot\,]$, i.e., $|\mco(t))=e^{\ii\mcl t}|\mco)$, similar to wave functions that evolve under the Hamiltonian $\mch$. Using the Liovillian superoperator, the autocorrelation function may be written as $C(t)=(\mco|e^{\ii\mcl t}|\mco)$. The Liouville space is equipped with an infinite-temperature inner product $(\mco_1|\mco_2)= \text{Tr}[\mco_1^\dagger \mco_2]$, which induces a norm via $||\mco||=\sqrt{(\mco|\mco)}$. \\
In the following, the central object of interest is the Liouvillian $\mathcal{L}$ represented in a particular basis $\{|\mathcal{O}_n)\}$, the so-called \textit{Krylov basis}. The Krylov basis is routinely constructed as part of the \textit{Lanczos algorithm}. In this basis, which is determined by some ``seed'' observable $\mco$,  the representation of the Liouvillian is tridiagonal. To initialize the algorithm, we take the normalized state $|\mco_0)=|\mco)$, i.e., $(\mco|\mco)=1$, and set $b_1=||\mcl \mco_0||$ as well as  $|\mco_1)=\mathcal{L}|\mco_0)/b_1$. Then we iteratively compute\\
\begin{align}
|\mcq_n)&=\mathcal{L}|\mco_{n-1})-b_{n-1}|\mco_{n-2})\,,\\\nonumber
b_n&=||\mcq_n||\,,\\\nonumber
|\mco_n)&=|\mcq_n)/b_n\,.
\end{align}
The tridiagonal representation of the Liouvillian in the \textit{Krylov basis} $\{|\mco_n)\}$ is then given by
\begin{equation}
\label{matrix}
L_{mn} =(\mco_m|\mcl|\mco_n)= \begin{pmatrix}
0 & b_1 \vphantom{\vdots} & 0 & ...\\
\vphantom{...}b_1 & 0\vphantom{\vdots} & b_2 & \\
0 & b_2 & 0 & \ddots\\
\vdots &  & \ddots & \ddots
\end{pmatrix}_{mn}\,,
\end{equation}
where the \textit{Lanczos coefficients} $b_n$ are real, positive numbers output by the algorithm. For our purposes, it is sufficient to assume a finite-dimensional space. Then, the algorithm halts at step $n=d+1$, where $d$ is the dimension of the Liouville space, and $L$ is a $d\times d$-matrix. Rewriting the Heisenberg equation of motion  in the Krylov basis yields
\begin{equation}
\label{eom}
\partial_t\varphi = -b_{n+1} \varphi_{n+1} + b_{n} \varphi_{n}\,,
\end{equation}
where we defined  $\varphi_n := \ii^{-n} (\mco_n|\mco(t))$. The initial condition is given by $\varphi_n(0)=\delta_{n0}$ and both $\varphi_n$ and $b_n$ are set to zero by convention when $n$ is ``out of bounds''.
The above Eq.\ \eqref{eom} takes the form of a discrete Schr\"odinger equation and can be numerically solved by familiar means of, e.g., exact diagonalization or iterative schemes like Runge-Kutta and Chebyshev polynomials.\\
As mentioned, the Lanczos coefficients $b_n$ can be interpreted as hopping amplitudes in a tight-binding model. Then, the correlation function coincides with the amplitude of the first site, i.e., $C(t)=\varphi_0(t)$.\\
For later reference, we introduce the spectral function $\Phi(\omega)$ as the Fourier transform of the correlation function, i.e.,
\begin{equation}
\Phi(\omega) = \int_{- \infty}^{\infty} e^{-\ii\omega t}\, C(t)\,\text{d}t\,.
\end{equation}
It can be shown that the Lanczos coefficients $b_n$ appear in the continued fraction expansion of $\Phi(\omega)$.
\begin{equation}
\Phi(\omega) = \text{Re } \dfrac{2}{\ii \omega + \dfrac{b_1^2}{\ii\omega+\dfrac{b_2^2}{\ii\omega+...}}}
\end{equation}
Consequently, there exists a (non-linear) one-to-one map between the Lanczos coefficients $b_n$ and the autocorrelation function $C(t)$. Thus, a set of $b_n$'s uniquely determines $C(t)$ and vice versa.\\
Lastly, we present the universal operator growth hypothesis as brought forth in Ref.\ \cite{berkley}. The hypothesis concerns the asymptotic behavior of the Lanczos coefficients $b_n$ and basically states that in generic, non-integrable systems the Lanczos coefficients of local, few-body observables grow asymptotically linear, i.e., above some $n$ the growth is given by
\begin{equation}
b_n \sim \alpha n + \gamma + o(1)\,,
\end{equation}
where $\alpha>0$ and $\gamma$ are real constants and $o(g_n)$ denotes some real sequence $f_n$ with $\lim_{n \rightarrow \infty} |f_n/g_n|=0$. In the special case of a one-dimensional system, the asymptotic growth is sub-linear due to an additional logarithmic correction \cite{berkley}.

\section{Probing for Stability}
\label{stabil}
\noindent
In this section, we devise our strategy to probe the stability of classes of dynamics with respect to certain types of perturbations. This section consists of the following parts:\
We present ways to find Lanczos coefficients corresponding to a chosen [class of] correlation functions [Sec.\ \ref{dlc}]. Next, we present four physically motivated requirements that the Lanczos coefficients should fulfill [Sec.\ \ref{require}]. Then, we discuss the particular type of perturbations we consider [Sec.\ \ref{pert}]. Finally, we introduce quantifiers that allow to evaluate the stability of the respective dynamics [Sec.\ \ref{detstab}].

\subsection{Dynamics and Lanczos coefficients}
\label{dlc}
\noindent
The preliminary section already touched on the relation between a correlation function $C(t)$ and the corresponding Lanczos coefficients $b_n$. As mentioned, this correspondence is one-to-one, thus, a set of $b_n$'s uniquely determines $C(t)$ and vice versa. However, the convergence can be quite subtle, i.e., there may be fairly similar dynamics with vastly different Lanczos coefficients. On the other side, similar Lanczos coefficients may lead to quite different dynamics. \\
In the numerical experiments below, we proceed as follows. First, we choose a correlation function $C'(t)$ whose stability we want to probe. We may specify $C'(t)$ as a concrete analytical function, or as a member of a class of functions, e.g., exponential decays. The exact Lanczos coefficients corresponding to $C'(t)$ are denoted by $b'_n$. These coefficients are, in general, unknown and obtaining them is quite a difficult task. Our objective is to find Lanczos coefficients $b_n$ corresponding to a correlation function $C(t)$ that should be practically indistinguishable from $C'(t)$ for all intents and purposes [$C(t) \simeq C'(t)$]. These ``approximate'' coefficients $b_n$ may be obtained in three different ways:\\
\textit{i}. The exact coefficients $b'_n$ may be analytically known. In this case $b_n=b'_n$ as well as $C(t)=C'(t)$.\\
\textit{ii}. The coefficients may be obtained via an educated guess. In this case, the $b_n$ may be quite different from $b'_n$ while still $C(t) \simeq C'(t)$.\\
\textit{iii}. The coefficients may be ``reverse-engineered'' from the correlation function $C'(t)$. For details see App.\ \ref{reveng}.\\
In the following, we will omit the technical distinction between $C(t)$ and $C'(t)$, since we assume that Lanczos coefficients can be found for which the two correlation functions are practically indistinguishable.

\subsection{Design of the Lanczos coefficients}
\label{require}
\noindent
In the following, we compare dynamics of correlation functions $C_A(t)$ and $C_B(t)$ with respect to their stability under a certain class of perturbations. Corresponding sets of Lanczos coefficients $b^A_n$ and $b^B_n$ may be obtained by one of the three methods mentioned in the last section. We demand that these Lanczos coefficients satisfy the following [partly physically motivated] requirements sufficiently well:\\
\textbf{\textit{i}}. The Lanczos coefficients $b^X_n$ should reproduce the respective correlation function $C_X(t)$ to an acceptable accuracy, where $X=A,B$ (this is sort of obvious and has already been addressed above).\\
\textbf{\textit{ii}}. The Lanczos coefficients $b^X_n$ should comply with the universal operator growth hypothesis, i.e., they should eventually attain linear growth.\\
\textbf{\textit{iii}}. The resulting correlation functions $C_A(t)$ and $C_B(t)$ should decay on more or less the same time scale. This is to ensure a fair comparison between the two, since faster dynamics are typically less affected by perturbations than slower ones.\\
\textbf{\textit{iv}}. The Lanczos coefficients $b^A_n$ and $b^B_n$ should be similar in magnitude. In particular, we demand that the quantity $\sum_n (b^X_n)^2$ is comparable for $X=A,B$. We show in App.\ \ref{appb} that this sum is related to the spectral variance of the Hamiltonian. Hence, this condition fosters the notion that the two correlation functions $C_A(t)$ and $C_B(t)$ originate from different observables while the underlying Hamiltonians are quite similar. In practice, a value of
\begin{equation}
q(b^\text{A}_n,b^\text{B}_n)=\dfrac{\sum_n (b^\text{A}_n)^2}{\sum_n (b^\text{B}_n)^2}
\end{equation}
close to unity is desirable.
\newpage

\subsection{Design of the perturbations}
\label{pert}
\noindent
The considered perturbations are designed on the level of Lanczos coefficients. The coefficients $b_n$ corresponding to some correlation function $C(t)$ will be slightly altered according to
\begin{equation}
\label{pert1}
\tilde{b}_n = b_n + \lambda v_n\,,
\end{equation}
where the perturbed coefficients are denoted by $\tilde{b}_n$. Here, $\lambda$ is the perturbation strength and $v_n$ is specified below. We show in \mbox{App.\ \ref{appc}} that this particular form of the perturbation yields a sensible scaling of the perturbed Hamiltonian with $\lambda$.\\
\noindent
In general, it is an intricate problem to determine how a perturbation in the form of $v_n$ corresponds to a perturbation $\mathcal{V}$ on the level of Hermitian matrices. This is an issue that we do not attempt to tackle in this paper. \\
Here, we try to make as few assumptions as possible and model the perturbation $v_n$ as random numbers [with the only restriction of a minimal correlation length, see below]. Concretely, we set
\begin{equation}
\label{pert2}
v_n = \sum_{k=1}^{N_\text{f}} x_k \cos(2 \pi n k /d) + y_k \sin(2 \pi n k /d)\,,
\end{equation}
where the $x_k, y_k$ are real, random numbers from a Gaussian distribution with zero mean and unit variance. They are normalized as $\sum_k x_k^2 + y_k^2 = 1$.\\
The sum in Eq.\ \eqref{pert2} is capped at a number $N_\text{f}$. This corresponds to excluding shorter wavelength or higher frequencies, which induces a minimal correlation length in the coefficients $\tilde{b}_n$ [reddish noise]. No truncation at all, i.e., $N_\text{f}=d$, would be equivalent to adding uncorrelated random numbers [white noise] to the coefficients $b_n$. This choice can be justified by invoking the tight-binding model interpretation in which the Lanczos coefficients represent hopping amplitudes between neighboring sites. Simply altering the hopping amplitudes in a random, uncorrelated manner leads to localization effects similar to Anderson localization \cite{anderson}. Through unsystematic varying of $N_\text{f}$, we find that the transition from unlocalized to to localized behavior is quite sharp. Since the aim is to study the stability of certain relaxation dynamics, we choose $N_\text{f}$ as large as possible, but small enough to avoid said localization effects. In practice, this amounts to $N_\text{f}\approx d/3 $. In Sec.\ \ref{patho}, we ease this restriction to identify ``pathological'' perturbations.\\
\newpage



\subsection{Assessing stability}
\label{detstab}
\noindent
Once the perturbation has acted on the coefficients $b_n$ and yielded the perturbed coefficients $\tilde{b}_n$, we can calculate the corresponding perturbed dynamics $\tilde{C}(t)$ as laid out in Sec.\ \ref{set}. Now, some tools are needed in order to judge how strongly the dynamics were altered by the perturbation. In particular, it is important to assess to what extent the perturbed dynamics still falls into the original class of functions to which the unperturbed dynamics belonged. For example, we may ask if $\tilde{C}(t)$ is still exponential, given that the original dynamics $C(t)$ was exponential [the decay constants may differ]. To this end, we attempt to fit the perturbed dynamics with a function that also described the unperturbed dynamics, e.g., we may try to fit $\tilde{C}(t)$ with $f(t)=A e^{-\mu t}$. In practice, the fit is obtained via a standard fitting routine.\\
Before we introduce quantifiers that measure the effect of the perturbation, some remarks on the fitting ansatz are in order.
Correlation functions are symmetric in time, thus, all odd moments necessarily vanish. In particular, correlation functions have zero slope at $t=0$. Further, we normalize the correlation function to unity at $t=0$. However, for fitting we choose functions that lack these features, i.e., the above exponential ansatz may yield an $A$ that slightly differs from unity. Further, the slope of an exponential at $t=0$ is never zero. This ``negligence'' is due to the fact that we are more interested in an overall description of relaxation dynamics, rather than in the details of the short-time behavior.\\
We assess the quality of a fit $f(t)$ by calculating ``how far off'' it is from the given perturbed dynamics. Concretely, we define a measure of the ``error'' or ``deviation'' $\epsilon$ by the expression
\begin{equation}
\label{error}
\epsilon = \sqrt{\dfrac{1}{N_\text{eq}}\sum_{n=0}^{N_\text{eq}} (\tilde{C}(t_n) - f(t_n))^2 }\,.
\end{equation} 
Here, the respective functions are evaluated at available points in time $t_n = n \delta t$, where $\delta t$ is the time step used to solve the equation of motion. The upper bound $N_\text{eq}$ corresponds to a time at which the dynamics in question has seemingly equilibrated \cite{note}. \\ 
To get a feeling of which numerical value of $\epsilon$ constitutes a good or bad fit, we refer to the exemplary numerical data displayed in Figs.\ \ref{exp3}, \ref{gauss3}, \ref{eex23} and \ref{geex23}.\\ 
We introduce a second quantifier $\sigma$ that measures how strongly the unperturbed dynamics are altered due to the perturbation in the first place.
\begin{equation}
\label{error2}
\sigma = \sqrt{\dfrac{1}{N_\text{eq}}\sum_{n=0}^{N_\text{eq}} (\tilde{C}(t_n) - C(t_n))^2 }\,.
\end{equation} 
The construction of this quantity is similar to the construction of the quantity $\epsilon$ that measures the fit quality.

\newpage

\section{Numerical analysis}
\label{numeric}
\noindent
In this section, we apply the presented strategy to certain relaxation dynamics. In particular, we investigate and compare the stability of two classes of dynamics. The first class in question is the class of damped oscillations, whose damping is due to an ordinary exponential factor. This class also includes simple exponential decays, which can be viewed as damped oscillations with zero frequency. These dynamics are ubiquitous in nature. For example, slow exponential dynamics may commonly arise whenever a system interacts weakly with an environment or whenever long-wavelength Fourier components  of  spatial densities  of  conserved  quantities are considered. Further, exponentially damped oscillations are routinely observed as short-wavelength components.\\
The possible choices for the second class of dynamics are of course manifold. Since an exhaustive analysis is impossible, we decide on a ``Gaussianized'' version of the first class. This is supposed to mean that the class includes Gaussian decays as well as oscillations damped by Gaussian factors.\\
The restriction of the second class of course stifles any aspiration of generality. Thus, the following results should be viewed as mere numerical experiments that corroborate the existence of stability in the first class.\\
Concretely,  in Sec.\ \ref{gaussexpo}, we compare exponential and Gaussian decay. In Sec.\ \ref{gausexpoo}, the comparison is between two damped oscillations, one with an exponential damping factor and the other with a Gaussian damping factor.
Lastly, in Sec.\ \ref{patho}, we identify ``pathological'' perturbations that destroy the considered dynamics.\\
Throughout the next sections, we set the dimension of the Liouville space to $d=10000$ and the frequency cut-off to $N_\text{f}=3333\approx d/3$. We repeat the strategy $N=1000$ times by drawing random numbers $x_k,y_k$ and present statistics for the quantifiers $\epsilon$ and $\sigma$.

\subsection{Exponential vs.\ Gaussian decay}
\label{gaussexpo}
\begin{figure}[t]
\includegraphics[width=\linewidth]{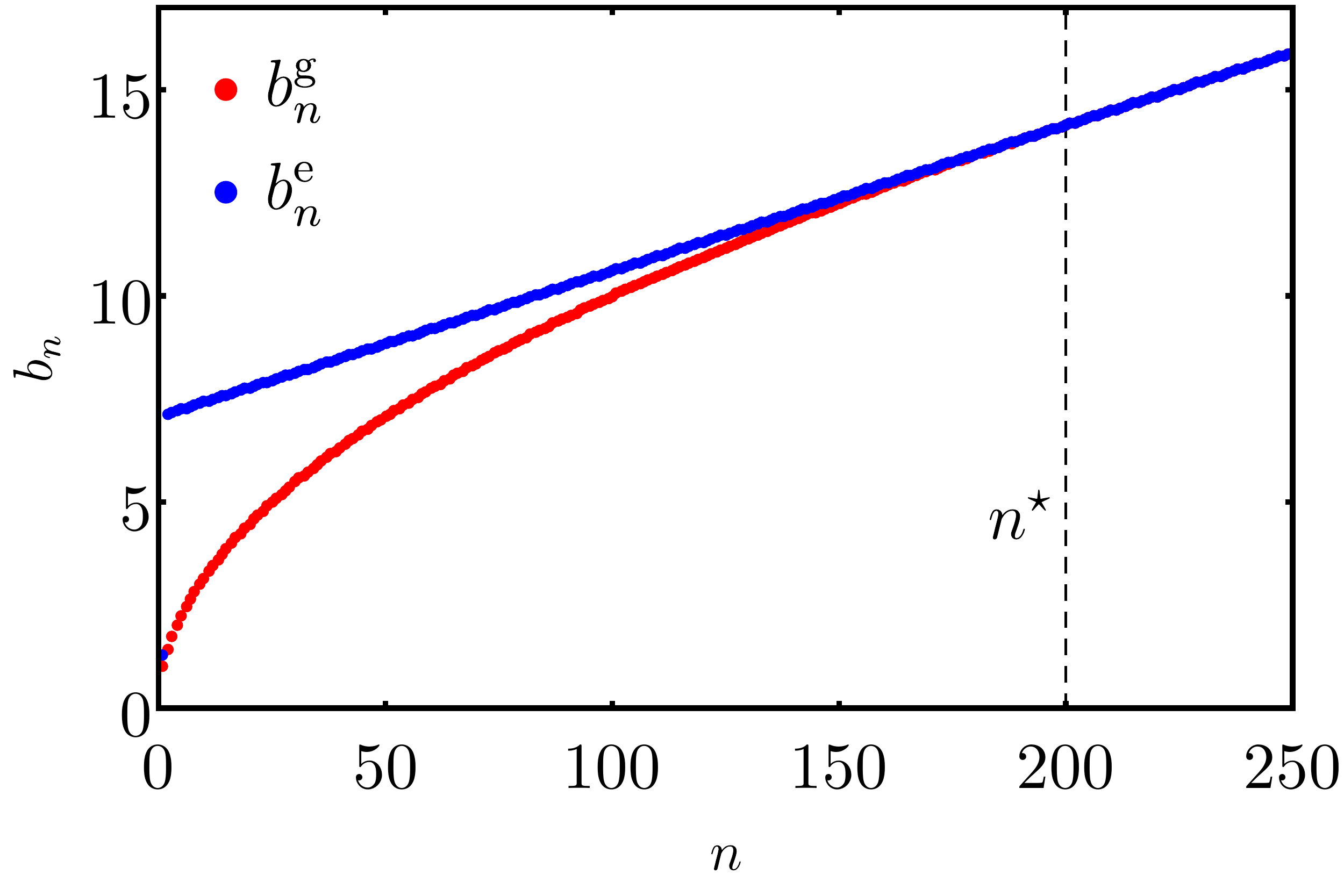}
  \caption{Lanczos coefficients corresponding to Gaussian decay (red) and exponential decay (blue). The square-root growth transitions to linear growth at $n=n^\star$.}
  \label{unpbn1}
  \end{figure}

\begin{figure}[t]
\includegraphics[width=0.97\linewidth]{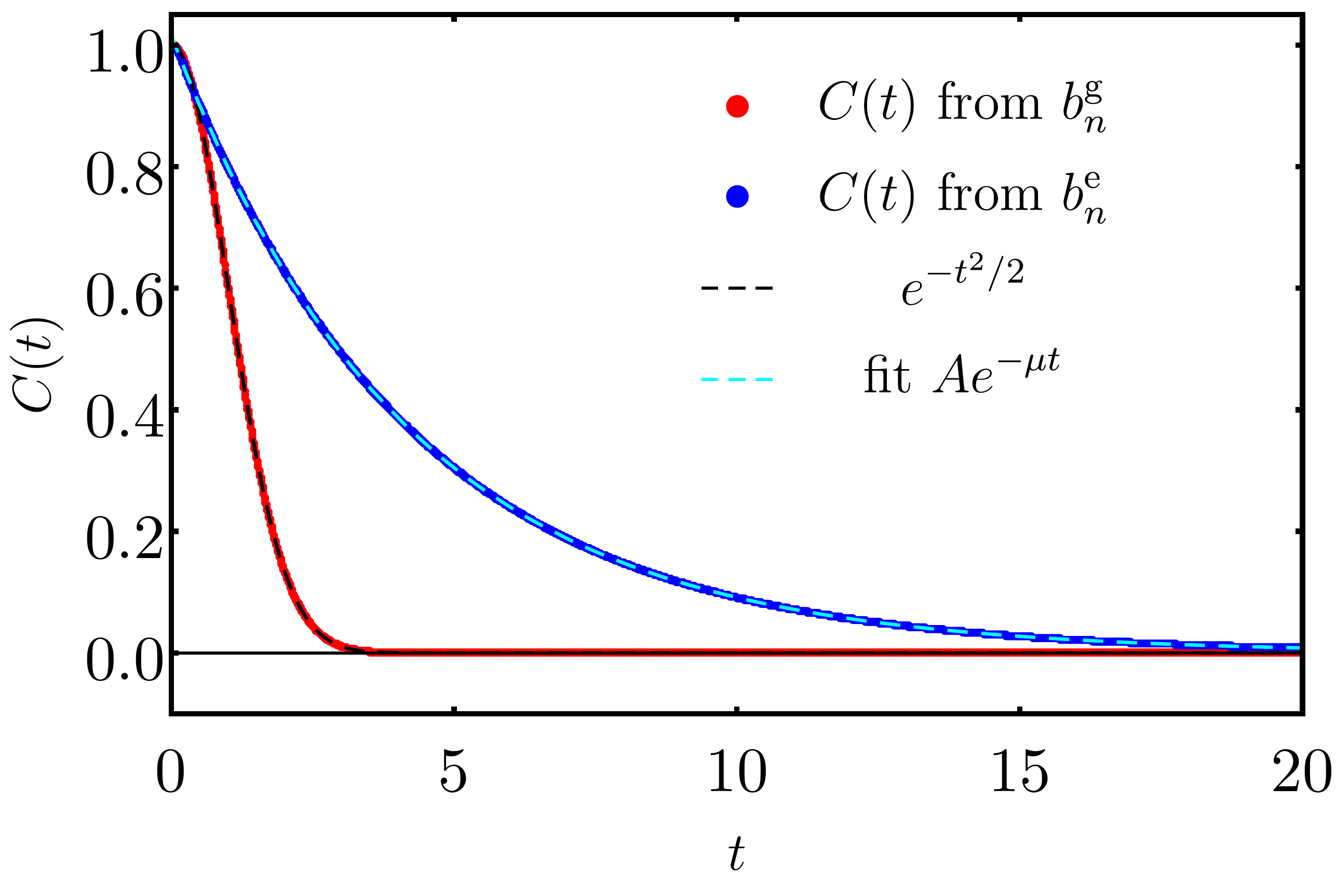}
  \caption{Unperturbed correlation functions $C(t)$ calculated from the respective Lanczos coefficients. Dashed curves indicate an exponential fit (cyan) 
  and an exact Gaussian $e^{-t^2/2}$ (black).}
  \label{unpdyn1}
    \end{figure}

\begin{figure}[b]
\includegraphics[width=0.97\linewidth]{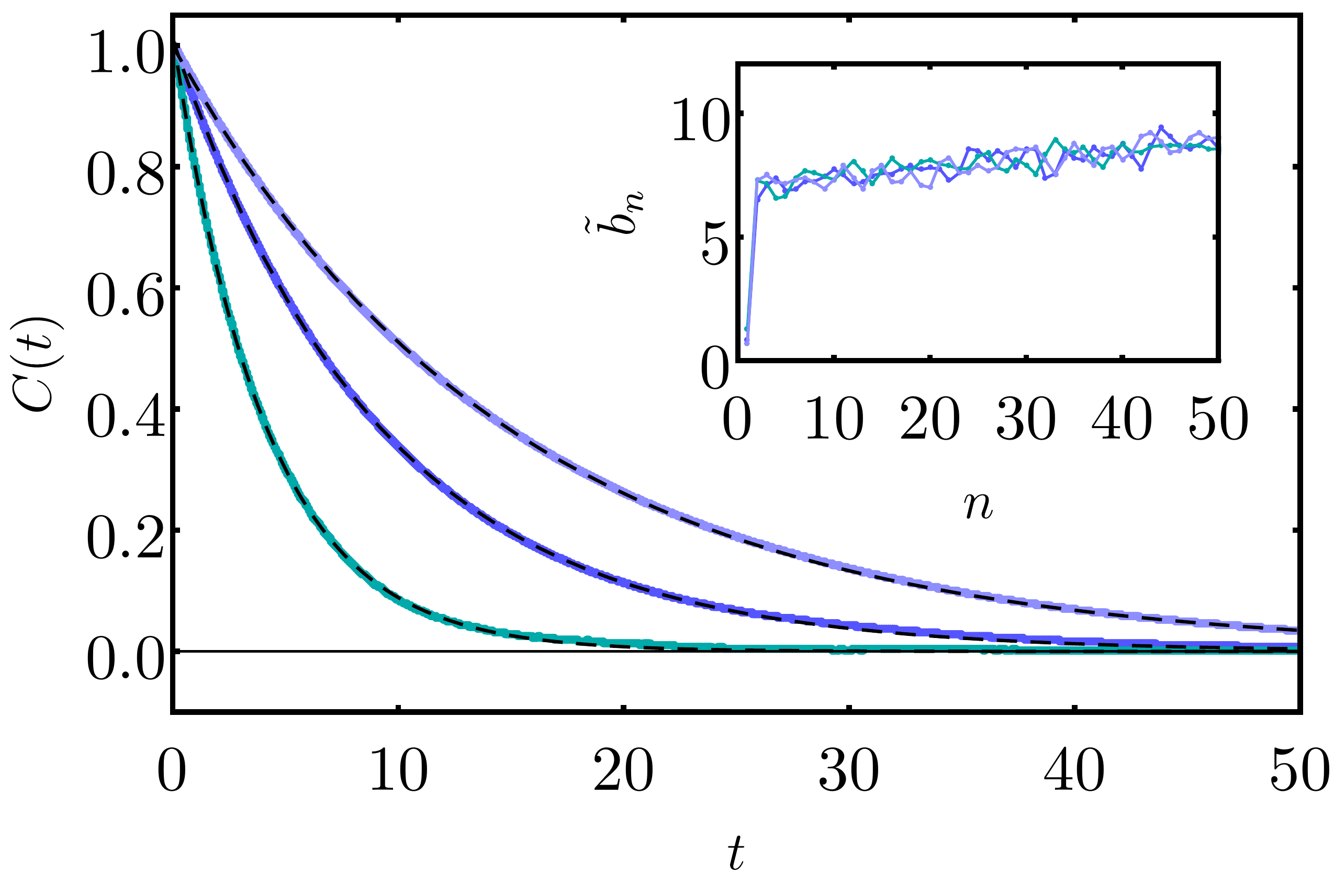}
  \caption{Three exemplary correlation functions originating from the perturbed exponential Lanczos coefficients $\tilde{b}_n^\text{e}$. Dashed, black lines indicate exponential fits. The quantifier $\epsilon$ of all three dynamics is relatively close to the mean value, see Fig. \ref{histo1}. Inset:\ corresponding perturbed Lanczos coefficients.}
  \label{exp3}
    \end{figure}

\begin{figure}[t]
\includegraphics[width=\linewidth]{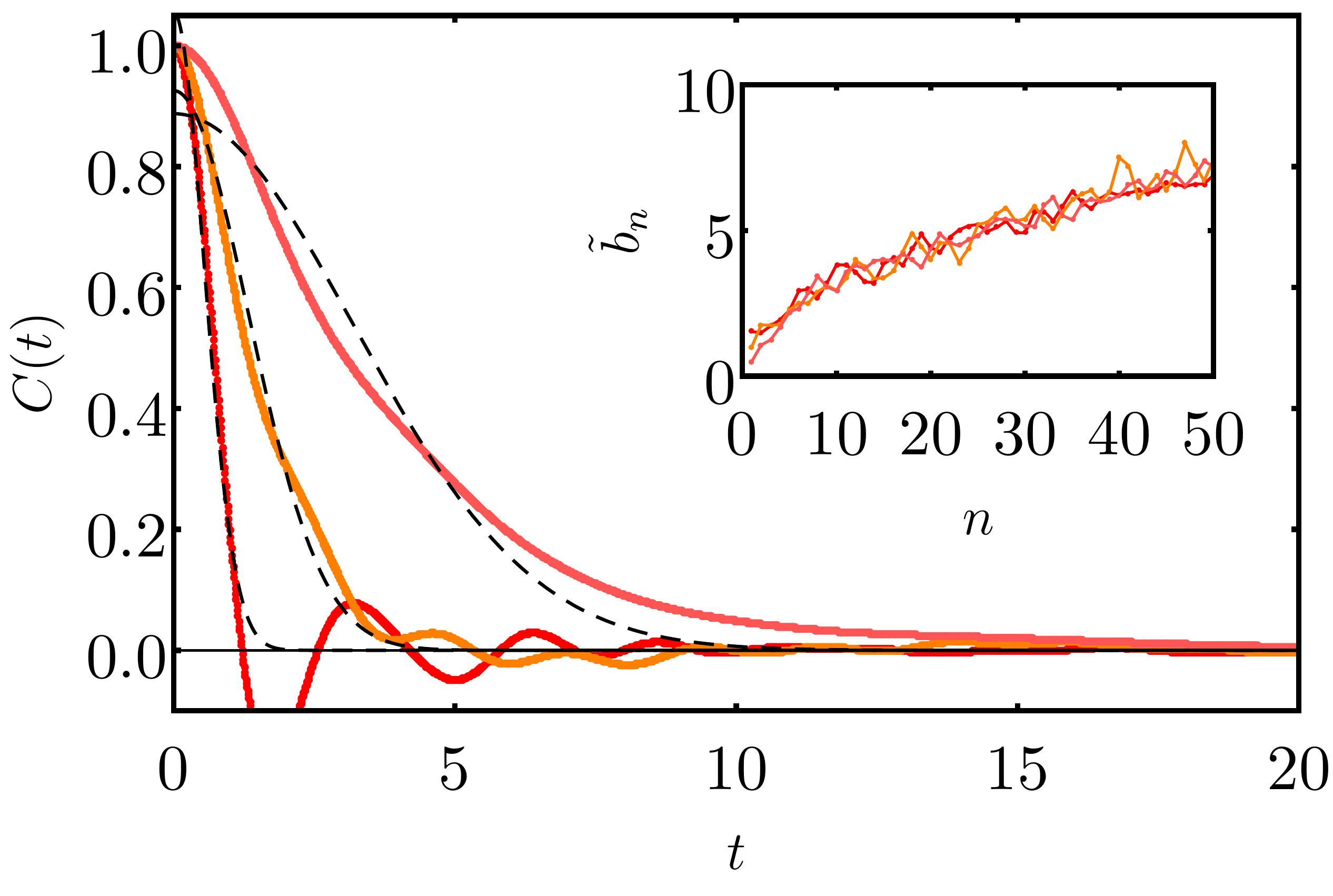}
  \caption{Three exemplary correlation functions originating from the perturbed Gaussian Lanczos coefficients $\tilde{b}_n^\text{g}$. Dashed, black lines indicate Gaussian fits. The quantifier $\epsilon$ of all three dynamics is relatively  close to the mean value, see Fig. \ref{histo1}. Inset:\ corresponding perturbed Lanczos coefficients.}
  \label{gauss3}
    \end{figure}

\noindent
In the first round of our stability investigation, an exponential decay competes against a Gaussian decay. Following the strategy laid out in the previous sections, the first step is to find suitable \lc that comply with the four conditions presented in Sec.\ \ref{require}. \\
We begin with the Gaussian decay. The Lanczos coefficients that exactly correspond to Gaussian decay of the form $C(t) = e^{-t^2/2}$ are analytically known [method one in Sec.\ \ref{dlc})], i.e., $b_n=\sqrt{n}$ \cite{gray}. Nevertheless, the second condition, the compliance with the operator growth hypothesis, needs to be satisfied. To this end, we choose a cutoff-point $n^\star$ at which the coefficients are continued in a linear fashion. This yields the following Lanczos coefficients [``g'' = ``Gaussian'']
\begin{equation}
\label{gauss}
b_n^\text{g}=
\begin{cases}
\sqrt{n}\,,\quad\quad\,\,\; n\leq n^\star\\
\alpha n + \gamma\,,\quad n> n^\star\,.
\end{cases}
\end{equation}
The parameters $\alpha$ and $\gamma$ are determined by the requirement of a smooth transition from square-root growth to linear growth, i.e., $\alpha=1/(2\sqrt{n^\star})$ and $\gamma= \sqrt{n^\star}/2$. The change from square-root to linear growth for $n>n^\star$ does not strongly affect the ``Gaussianity'' of the correlation dynamics [compare Fig.\ \ref{unpdyn1}] such that condition one remains fulfilled.\\
Next, coefficients for an (approximate) exponential decay need to be found. As mentioned, a correlation function can never be truly exponential, since it necessarily features zero slope at $t=0$. Thus, we only require the exponential decay to be present after a short Zeno-time, which is usually exceedingly short compared to the relaxation time  \cite{zeno1,zeno2}. We achieve an approximate exponential decay via an educated guess [method two in \mbox{Sec.\ \ref{dlc}}]. Slow dynamics are characterized by a relatively small first coefficient $b_1$, followed by a jump to a larger remainder of coefficients $b_{n \geq 2}$. Thus, we make the following ansatz for the coefficients [``e'' = ``exponential'']
\begin{equation}
\label{expo}
b_n^\text{e}=
\begin{cases}
a\,,\quad\quad\quad\;\; n= 1 \\
\alpha n + \gamma\,,\quad n \geq 2 \,.
\end{cases}
\end{equation}
The parameter $a$ is set to $1.2$, a similar magnitude as the first Gaussian coefficient $b_1^\text{g}$. The other parameters $\alpha$ and $\gamma$ are the same as in the Gaussian case. Thus, condition two, i.e., compliance with the universal operator growth hypothesis, is fulfilled. The Lanczos coefficients as defined in Eq.\ \eqref{gauss} and Eq.\ \eqref{expo} are depicted in Fig.\ \ref{unpbn1}. Since both sets of coefficients coincide for $n>n^\star$, condition four is satisfied as the quantity $q(b^\text{g}_n,b^\text{e}_n)= 0.999993$
is sufficiently close to unity.
  The corresponding dynamics are depicted in Fig.\ \ref{unpdyn1}. The ansatz for the coefficients $b^\text{e}_n$ in Eq.\ \eqref{expo} indeed yields a nice exponential decay. A fit of the form $A e^{-\lambda t}$ with $A=1.02$ and $\lambda= 0.24$ captures the dynamics quite well. Thus, condition one is sufficiently fulfilled for the exponential decay as well. \\
Further, we can extract from Fig.\ \ref{unpdyn1} that both dynamics decay on more or less the same time scale. Therefore, we also view condition three as satisfied. If any, the Gaussian dynamics is faster and thus less prone to perturbations.\\
Now that we have checked all four conditions presented in Sec.\ \ref{require}, we are ready to apply the perturbation as laid out in Sec. \ref{pert}. We set $\lambda=0.5$. Three exemplary perturbed dynamics with respective fits for the exponential case are depicted in \mbox{Fig.\ \ref{exp3}}. \\
 It is evident that the perturbed dynamics are still nicely described by an exponential, only the decay constant changes. For the Gaussian decay, three exemplary perturbed dynamics with respective fits are depicted in \mbox{Fig.\ \ref{gauss3}}. Two displayed perturbed Gaussian curves feature oscillations, which can not possibly be captured by a Gaussian fit ansatz. The three depicted fits are much worse than in the exponential case. Insets show corresponding Lanczos coefficients.

    \newpage

\begin{figure}[t]
\includegraphics[width=\linewidth]{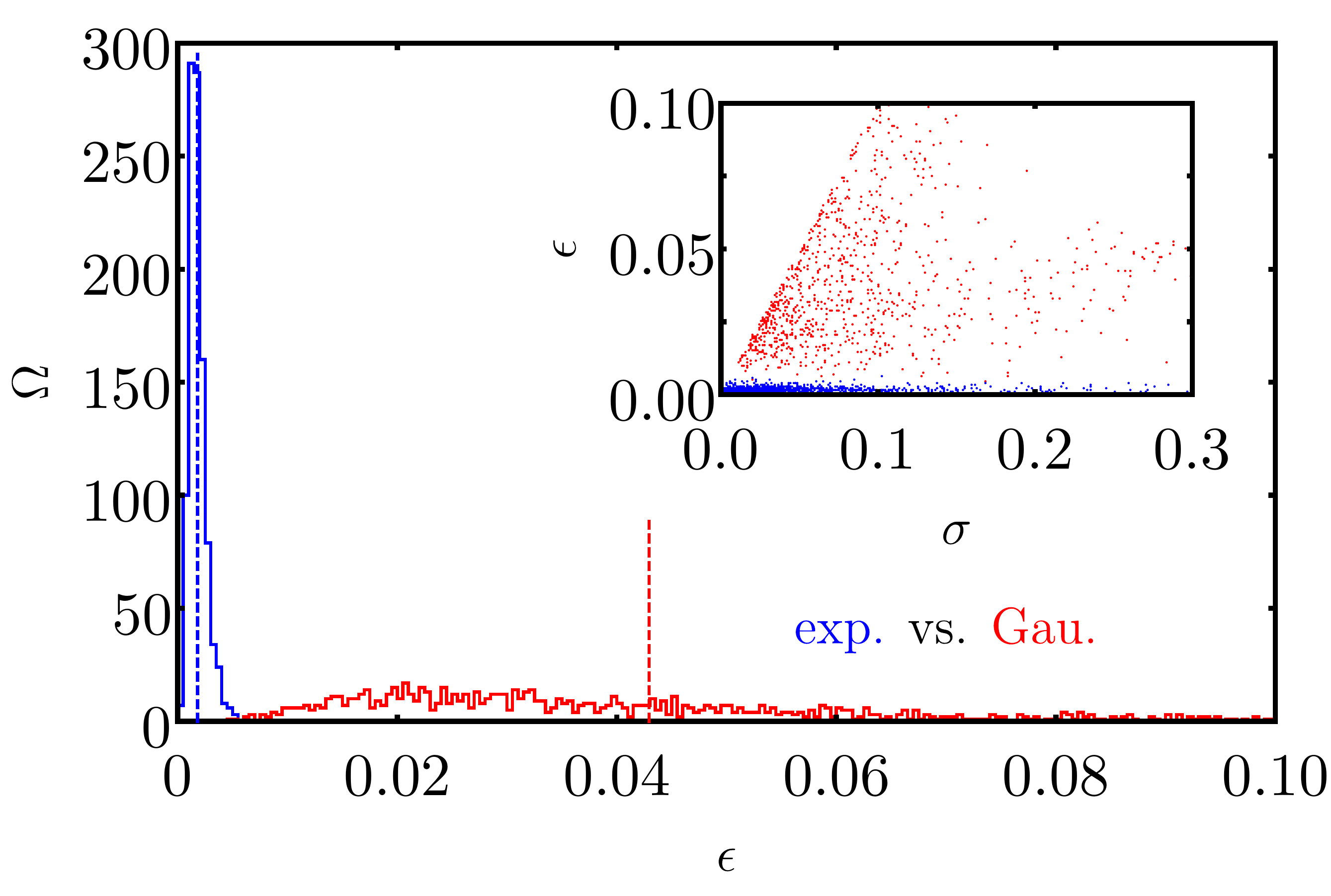}
  \caption{Histogram of the fit quality measure $\epsilon$ with a bin size of $5\cdot 10^{-4}$. Dashed lines indicate respective mean values. The stability of the exponential decay is evident. In contrast, the Gaussian decay does not seem to be stable, as the deviations $\epsilon$ become quite large.
  Inset:\ scatter plot of all points $(\sigma_i,\epsilon_i)$. Both dynamics are equally affected by the perturbation.}
  \label{histo1}
    \end{figure}

\noindent
 The impression from these six exemplary curves is confirmed in Fig.\ \ref{histo1}, which shows histograms of deviations $\epsilon$ of the fits from the respective perturbed dynamics. The symbol $\Omega$ denotes the number of values $\epsilon$ within a bin of size $5\cdot 10^{-4}$.  There is a clear division between the exponential cases (blue) and the Gaussian cases (red).\\
 The deviations $\epsilon$ of the exponential decays are much smaller than those of the Gaussian decays. In particular, the mean value $\bar{\epsilon}_\text{e}=0.002$ (indicated by a blue, dashed line) in the exponential case is about twenty times smaller than in the Gaussian case $\bar{\epsilon}_\text{g}=0.042$. For a visualization of this comparison see Fig.\ \ref{exp3} and Fig.\ \ref{gauss3}, whose exemplary curves feature values of $\epsilon$ close to the respective mean values. \\
We have to be aware that this apparent stability of the exponential may be due to the possibility that exponential decays are generally less affected by our constructed perturbation than the Gaussian decays. 
To check this possibility and to show that this is not the case, we consider the quantifier $\sigma$, which measures the difference between the perturbed and the unperturbed dynamics. The inset of Fig.\ \ref{histo1} depicts a scatter plot of all $1000$ pairs $(\sigma_i,\epsilon_i)$. If one decay would be consistently less affected than the other, a cluster of points to the left [small values of $\sigma$] would emerge.
This is evidently not the case, as the distribution along the horizontal axis is relatively similar for both decays. The existence of the diagonal edge is expected,  since there can be no fit that is ``farther away''  from the perturbed dynamics than the unperturbed dynamics itself, which is, of course, also an eligible candidate for fitting.\\
Thus, we conclude that exponential decays is indeed stable with respect to perturbations [as in Eq.\ \eqref{pert2}]. In contrast, Gaussian decays seem to be quite unstable. This is the first main result of the paper at hand.

    \newpage
    
\subsection{Damped oscillations:\ exponential vs.\ Gaussian}
\label{gausexpoo}
\noindent
In the second round, we investigate and compare two types of damped oscillations, one with an exponential damping factor, the other with a Gaussian damping factor. As in the last section, first, two sets of Lanczos coefficients need to be found that satisfy all four conditions imposed in Sec.\ \ref{require}.\\
 We again begin with the Gaussian case.
The coefficients $b_n^\text{gdo}$ [``gdo'' $=$ ``Gaussian damped oscillation''] that correspond to an oscillation damped by a Gaussian factor a neither analytically known, nor can we make an educated guess. This only leaves the third method, in which we ``reverse-engineer'' the coefficients from the dynamics itself. To this end, we choose a particular correlation function $C(t)=e^{-t^2/8}\cos(2t)$ and proceed as detailed in App.\ \ref{reveng}. This procedure allows us to obtain about $50$ Lanczos coefficients. After that, the coefficients are continued ``by hand'' in a manner that first, respects the pattern exhibited by the coefficients so far, and second, eventually becomes linear.\\
For the exponentially damped oscillation we can again make an educated guess [method two] for the coefficients $b_n^\text{edo}$ [``edo'' $=$ ``exponentially damped oscillation''].\\ We set the first two coefficients $b_1,b_2$ to some values ($b_1=2.0$, $b_2=1.6$), followed by a jump to larger coefficients $b_{n\geq 3}$, which then grow in a linear fashion. The slope is determined by and coincides with the slope in the Gaussian case.\newpage
            \begin{figure}[t]
\includegraphics[width=\linewidth]{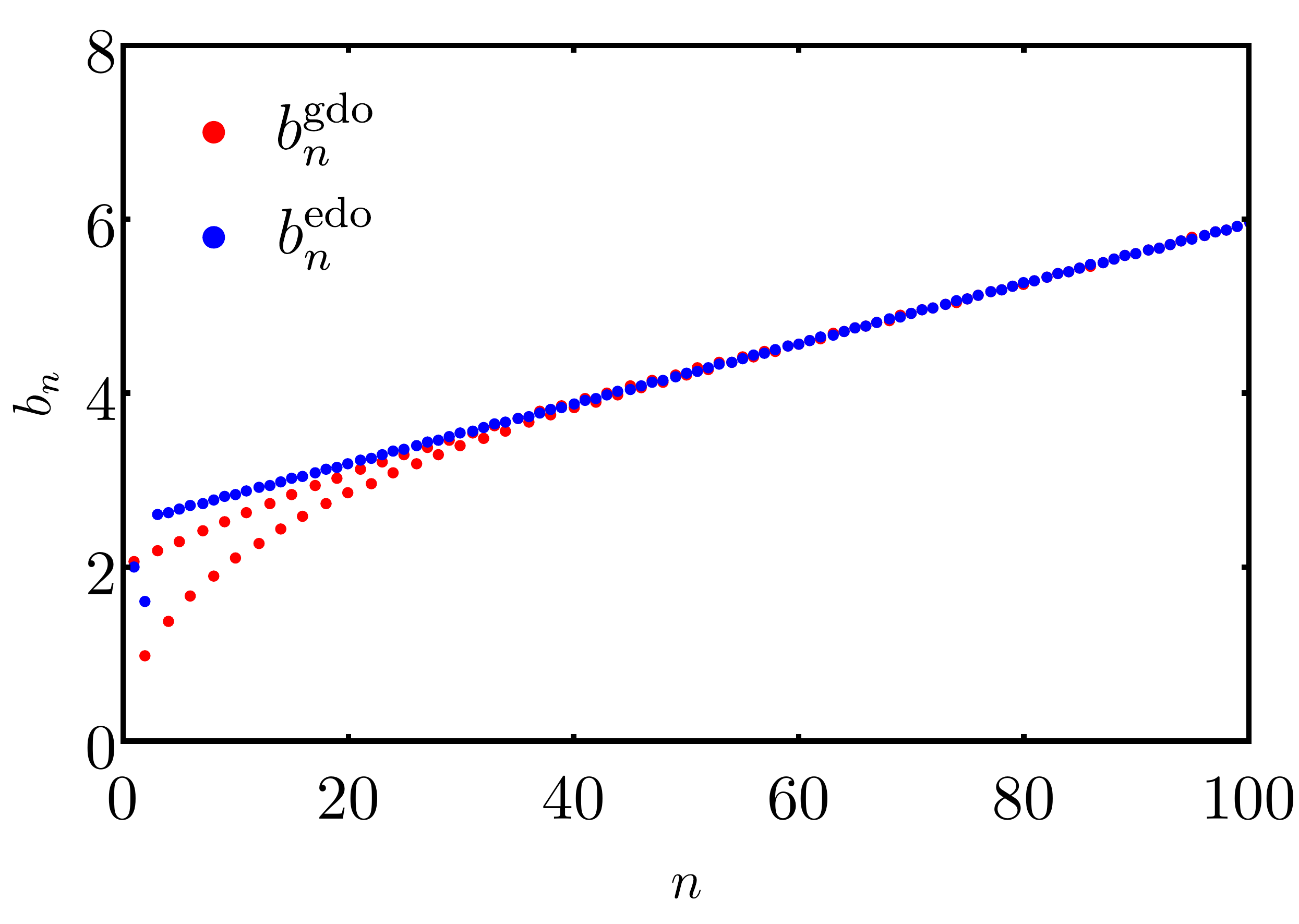}
  \caption{Lanczos coefficients corresponding to an oscillation damped by a Gaussian (red) and an oscillation damped by an exponential (blue).}
  \label{coeff2}
    \end{figure} 
    
        \begin{figure}[b]
\includegraphics[width=\linewidth]{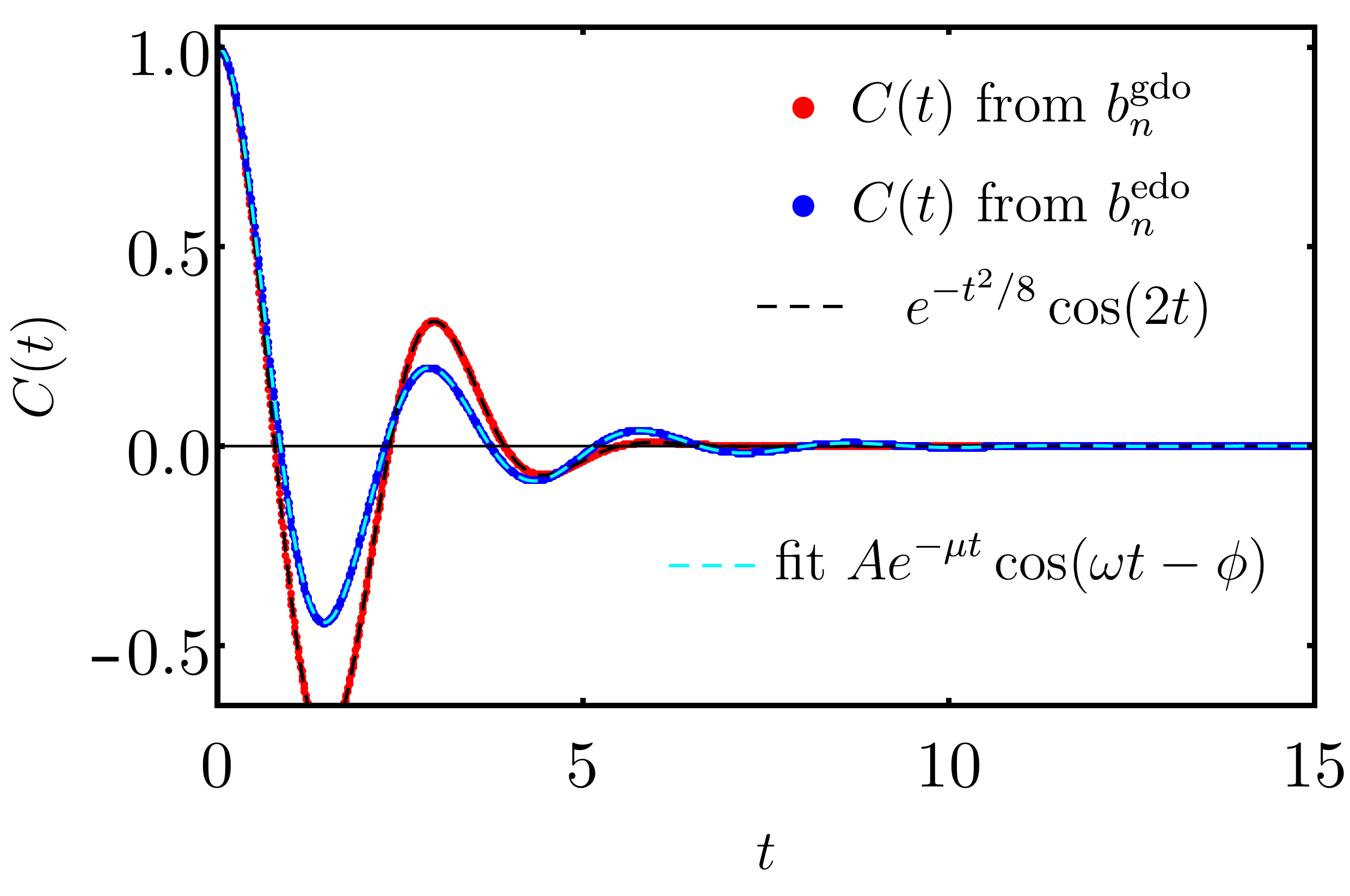}
  \caption{Unperturbed correlation functions $C(t)$ calculated from the respective Lanczos coefficients. Dashed curves indicate a fit of an exponentially damped oscillation (cyan) and the oscillation damped by a Gaussian $e^{-t^2/8}\cos(2t)$ (black).}
  \label{dyn2}
    \end{figure}     
    
    \noindent
The unperturbed Lanczos coefficients and corresponding correlation functions are depicted in Fig.\ \ref{coeff2} and \mbox{Fig.\ \ref{dyn2}}, respectively. The ``reverse-engineered and continued by hand'' coefficients $b_n^\text{gdo}$ indeed reproduce the given correlation function $C(t)=e^{-t^2/8}\cos(2t)$. Further, the guess for $b_n^\text{edo}$ yields a nice exponentially damped curve, i.e., the fit ansatz with [$A=1.04$, $\mu=0.57$, $\omega=0.32$, $\phi=2.19$] nicely captures the dynamics. Therefore, condition one is satisfied for both cases. The two sets of coefficients were designed to eventually attain linear growth, thus condition two is fulfilled. It is evident from Fig.\ \ref{dyn2} that both dynamics decay on similar time scales, rendering condition three satisfied. Lastly, the quantity $q(b_n^\text{gdo},b_n^\text{edo})$ practically  equals unity. Thus, all four conditions are met.\newpage

\begin{figure}[t]
\includegraphics[width=\linewidth]{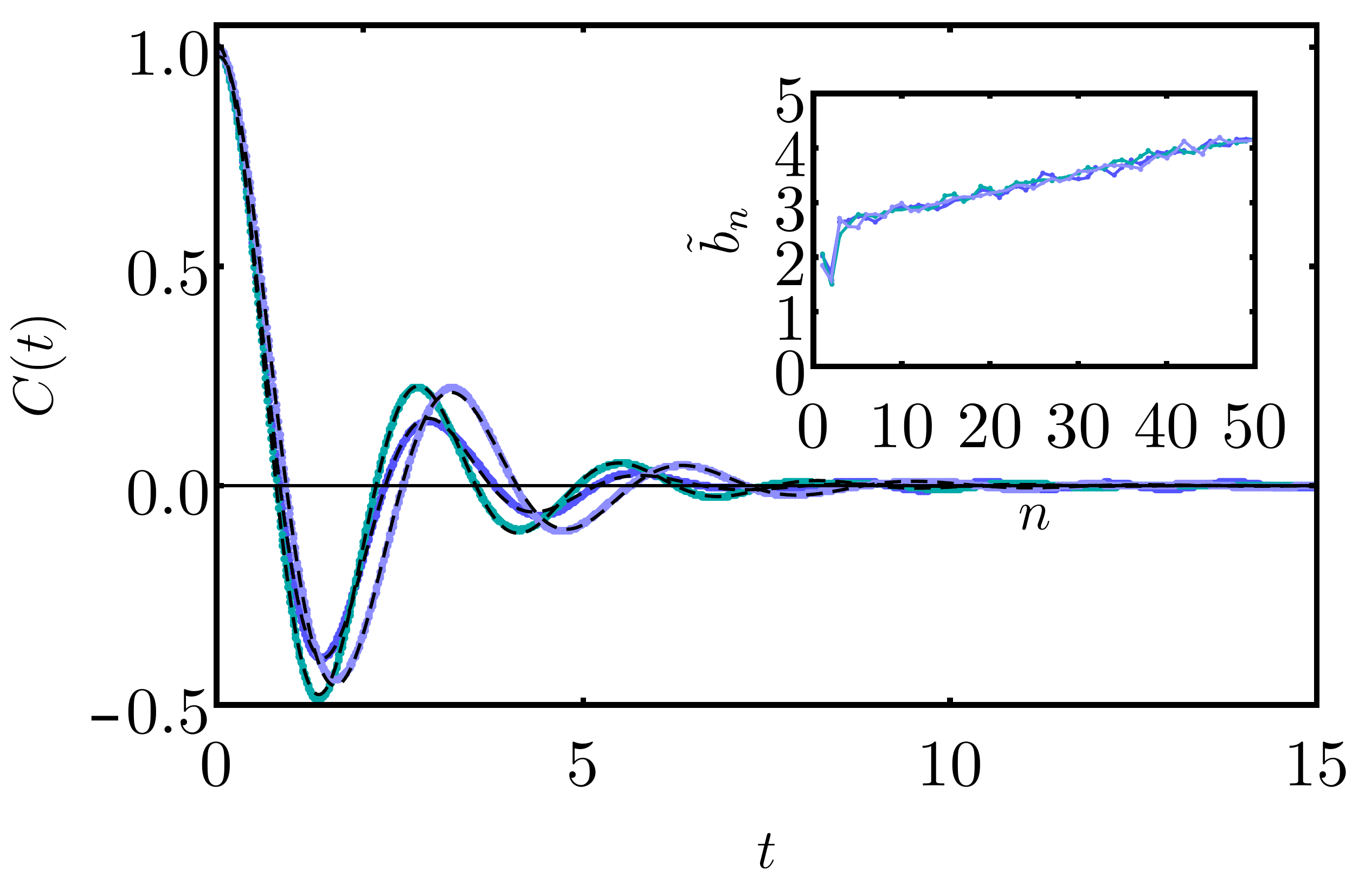}
  \caption{Three exemplary correlation functions originating from the perturbed Lanczos coefficients $\tilde{b}_n^\text{edo}$. Dashed, black lines indicate $4$-parametric fits of exponentially damped oscillations. The quantifier $\epsilon$ of all three dynamics is relatively close to the mean value, see Fig.\ \ref{histo2}. The perturbation not only changes the decay constant, but also the frequency. Inset:\ corresponding perturbed Lanczos coefficients.}
  \label{eex23}
    \end{figure}

\begin{figure}[b]
\includegraphics[width=\linewidth]{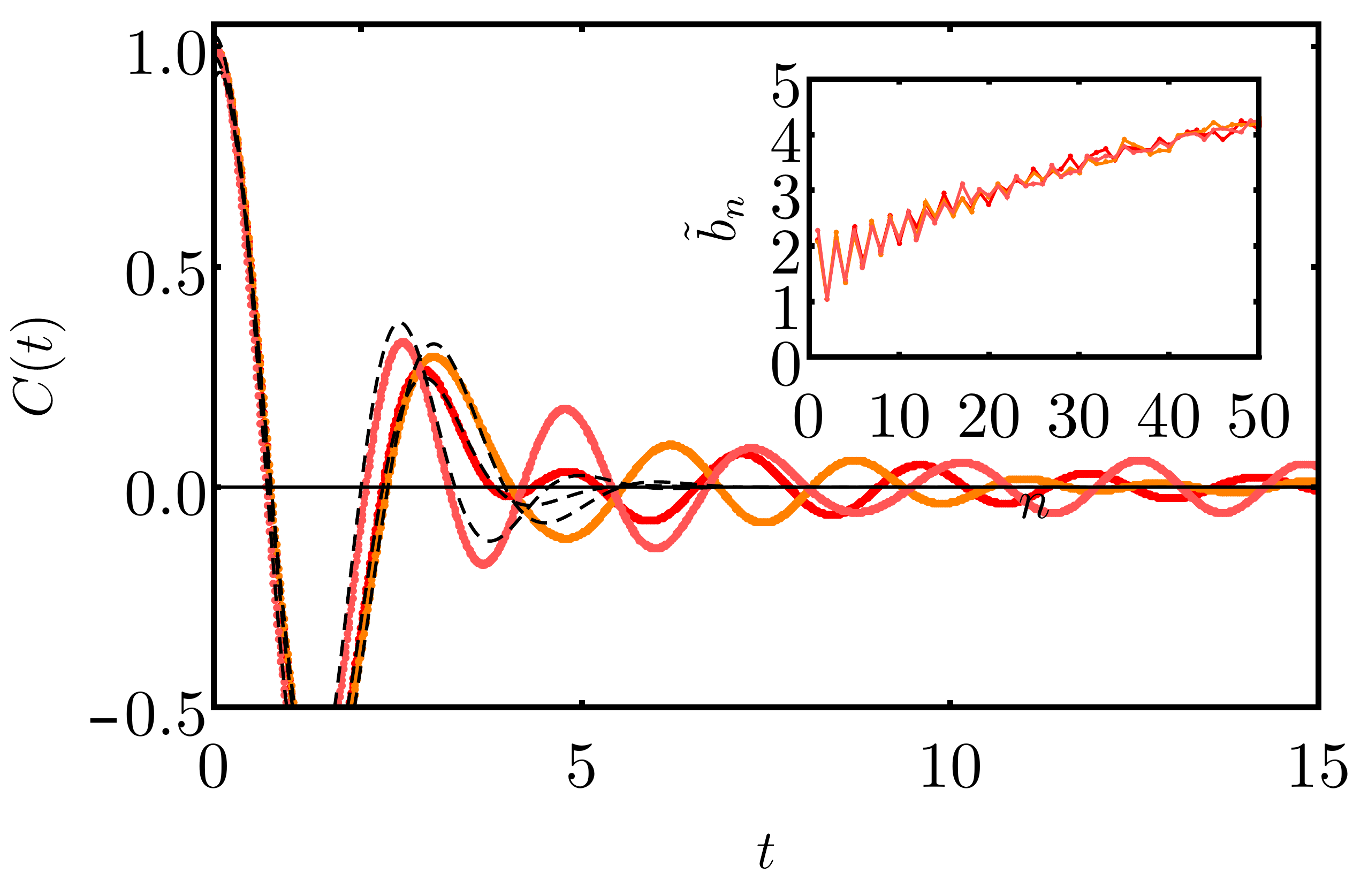}
  \caption{Three exemplary correlation functions originating from the perturbed Lanczos coefficients $\tilde{b}_n^\text{gdo}$. Dashed, black lines indicate $4$-parametric fits of oscillations damped by a Gaussian. The quantifier $\epsilon$ of all three dynamics is relatively close to the mean value, see Fig.\ \ref{histo2}. Inset:\ corresponding perturbed Lanczos coefficients.}
  \label{geex23}
    \end{figure} 

\noindent
Next, we switch on the perturbation and set $\lambda=0.1$. Exemplary perturbed dynamics are displayed in Fig.\ \ref{eex23} and Fig.\ \ref{geex23}, respectively. The perturbed dynamics are fitted with the $4$-parametric ansatz $C(t)=Ae^{-\mu t}\cos(\omega t -\phi)$ in the exponential and $C(t)=Ae^{-\mu t^2}\cos(\omega t -\phi)$ in the Gaussian case.
While the perturbed coefficients in the exponential case still lead to correlation functions that are within the class of exponentially damped oscillations [with different decay constants $\mu$ and frequencies $\omega$], the same can not be said for the Gaussian case, where the perturbed dynamics decay too slowly and can not by captured by the fit ansatz above.\newpage

\begin{figure}[t]
\includegraphics[width=\linewidth]{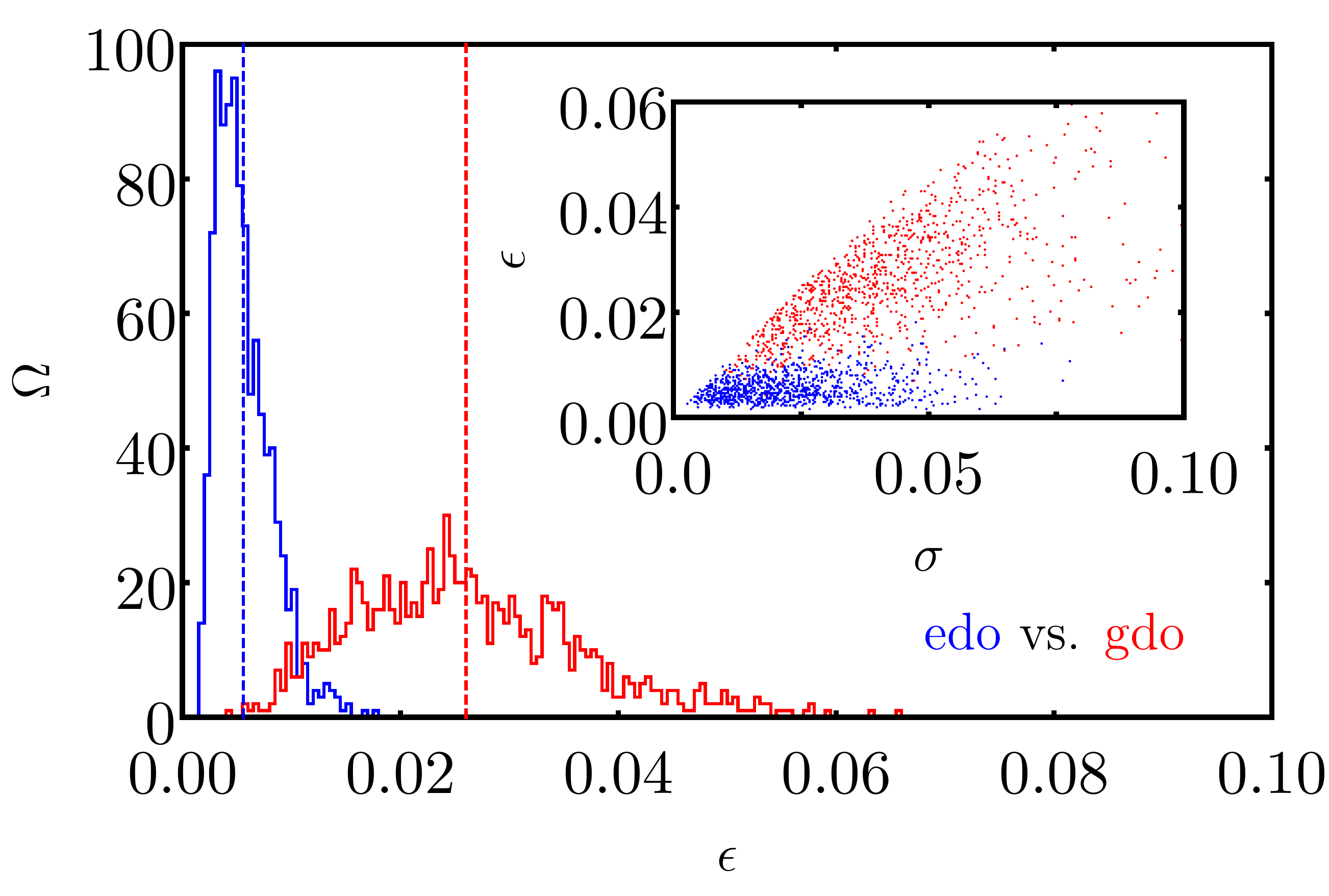}
  \caption{Histogram of the fit quality measure $\epsilon$ with a bin size of $5\cdot 10^{-4}$. Dashed lines indicate respective mean values. The stability of the exponentially damped oscillations is evident. In contrast, the Gaussian counterpart does not seem to be stable, as the deviations $\epsilon$ become quite large.\\
   Inset:\ scatter plot of all points $(\sigma_i,\epsilon_i)$. Both dynamics are more or less equally affected by the perturbation.}
  \label{histo2}
    \end{figure} 

\noindent
The impression from the six exemplary curves is confirmed in \mbox{Fig.\ \ref{histo2}}, which displays histograms of the quantifier $\epsilon$.  Again, the division into two distinct peaks is evident. The respective mean values $\overline{\epsilon}_\text{gdo} = 0.026$ and $\overline{\epsilon}_\text{edo} = 0.006$ are indicated as dashed, vertical lines and differ by a factor of about five. For a visualization of what these values mean for the fit quality, see Fig.\ \ref{eex23}
 and Fig.\ \ref{geex23}, whose curves feature values of $\epsilon$ close to the respective mean values.
Thus, we conclude that the exponentially damped dynamics are quite stable and, in particular, more stable than the oscillations damped by a Gaussian. The disparity between the two is not as strong as for the decays in the previous section [however, different values of $\lambda$ may not be directly comparable.]\\
The inset of Fig.\ \ref{histo2} again shows a scatter plot of all points  $(\sigma_i,\epsilon_i)$. The blue cluster of dots is a little more concentrated to small values of $\sigma$, indicating that the exponentially damped oscillations are a little less altered by the perturbation than their Gaussian counterpart. However, the marginal distribution over $\sigma$ is still less partitioned into two peaks than the one over $\epsilon$ [which is the data in the histogram itself]. Hence, the stability of exponentially damped oscillations is still due to the nature of the dynamics and not due to a smaller effect of the perturbation on the dynamics.\\
We conclude that exponentially damped oscillations are stable with respect to perturbations [as in Eq.\ \eqref{pert2}]. In contrast, oscillations damped by a Gaussian seem to be quite unstable. This is the second main result of the paper at hand.
\newpage

\begin{figure}[t]
\includegraphics[width=\linewidth]{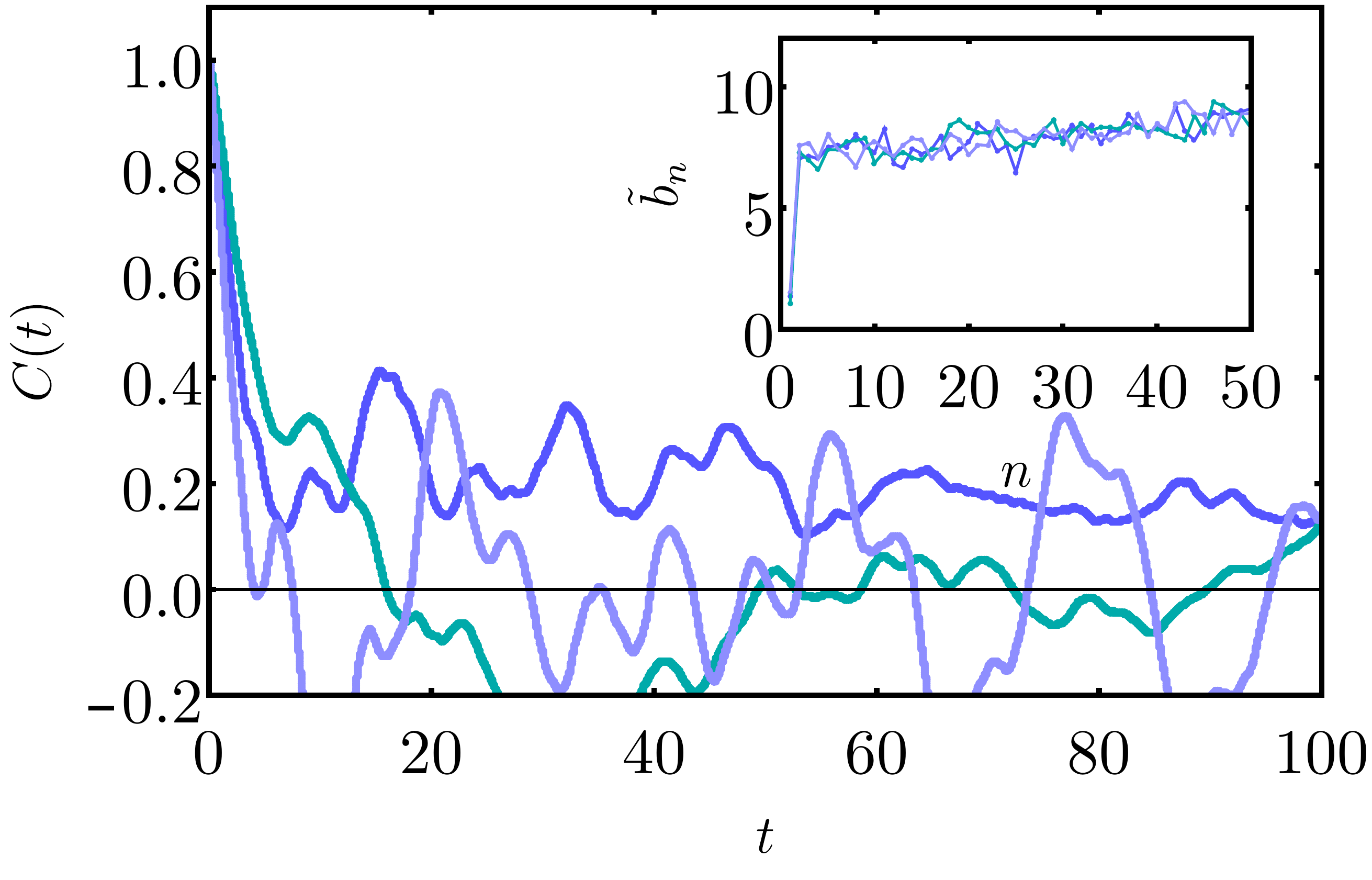}
  \caption{Three exemplary dynamics originating from perturbed Lanczos coefficients $\tilde{b}^\text{e}_n$. The perturbation consists of uncorrelated random numbers, i.e., $N_\text{f}=d$ in Eq.\ \eqref{pert2}. The exponential decay is completely destroyed and the correlation function does not seem to equilibrate at all. Inset:\ corresponding perturbed Lanczos coefficients.}
  \label{allf}
    \end{figure}

\subsection{Pathological perturbations}
\label{patho}
\begin{figure}[t]
\includegraphics[width=\linewidth]{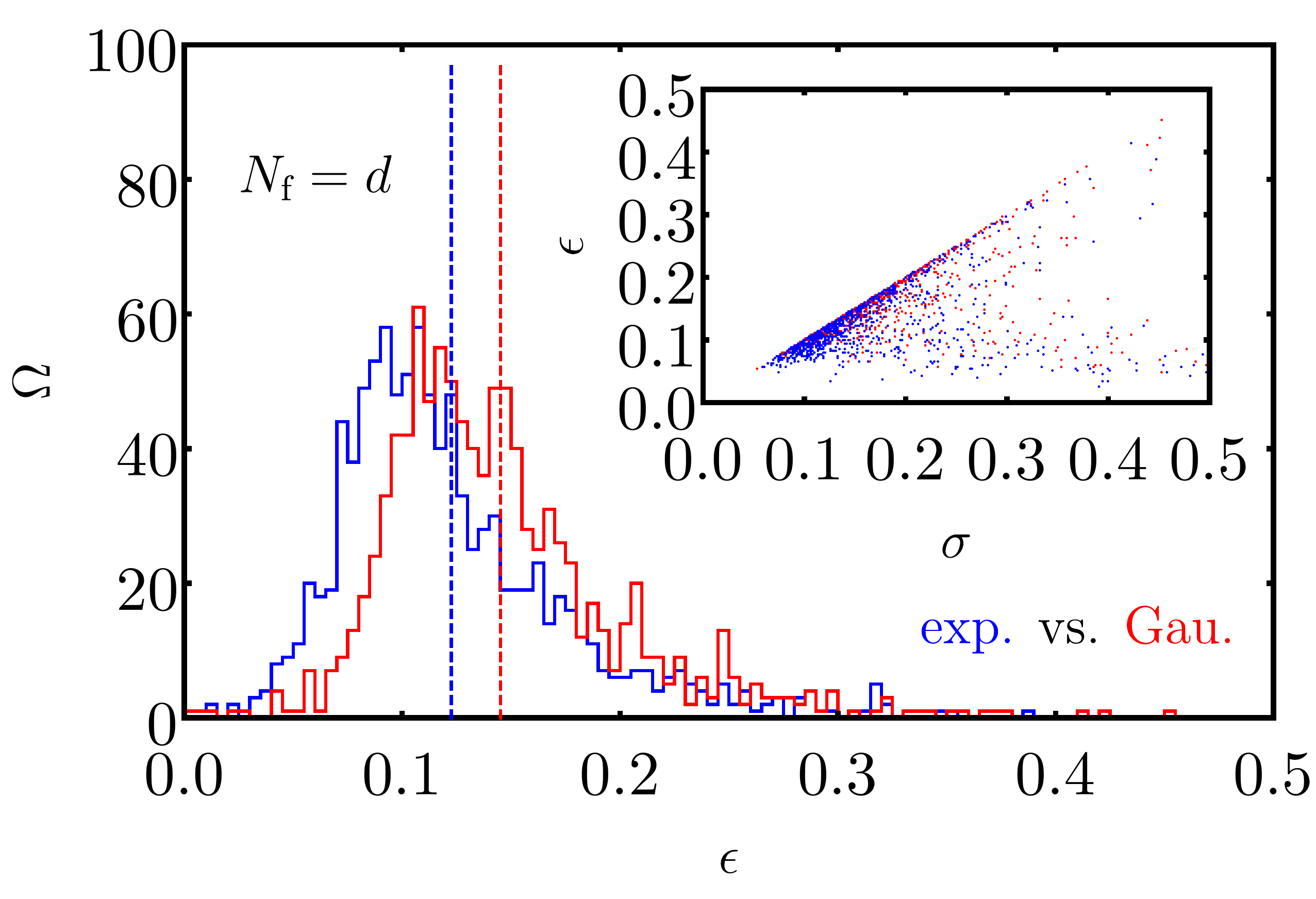}
  \caption{Histogram of the fit quality measure $\epsilon$ with a bin size of $5\cdot 10^{-3}$. Dashed lines indicate respective mean values. For both oscillating cases the deviations $\epsilon$ are quite large. Thus, both relaxation dynamics are unstable with respect to the perturbation including all wave lengths.
   Inset:\ scatter plot of all points $(\sigma_i,\epsilon_i)$. Points accumulate along the diagonal edge indicating that the respective dynamics are heavily altered due to the perturbation.}
  \label{histoall1}
    \end{figure}

\begin{figure}[t]
\includegraphics[width=\linewidth]{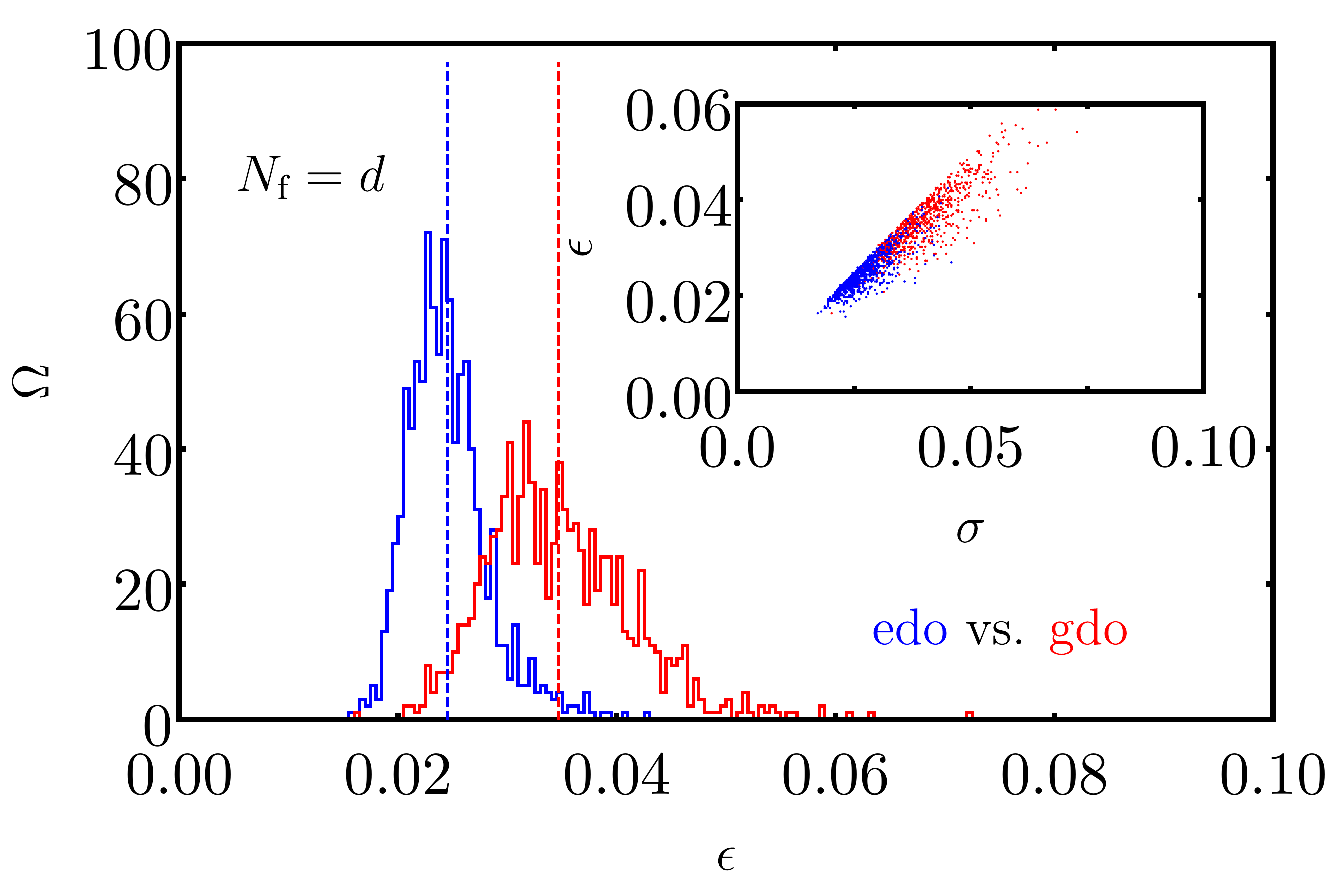}
  \caption{Histogram of the fit quality measure $\epsilon$ with a bin size of $5\cdot 10^{-4}$. Dashed lines indicate respective mean values. For both the exponential and the Gaussian case the deviations $\epsilon$ are extremely large. Thus, both relaxation dynamics are unstable with respect to the perturbation including all wave lengths.
   Inset:\ scatter plot of all points $(\sigma_i,\epsilon_i)$. Points accumulate along the diagonal edge indicating that the respective dynamics are heavily altered due to the perturbation.}
  \label{histoall2}
    \end{figure}

\noindent
In this section, we lift the restriction imposed on the $v_n$ earlier, which was that frequencies were cut at $N_\text{f}\approx d/3$. Instead, we include all frequencies in the construction of the perturbation, which amounts to setting $N_\text{f}=d$. This case simply corresponds to adding uncorrelated random numbers to the unperturbed coefficients $b_n$. Otherwise, the strategy is pursued in the same manner as above.\\
Three exemplary dynamics for exponential decays are depicted in Fig.\ \ref{allf} [for conciseness, we refrain from showing more exemplary data for the other cases]. As is evident, the exponential decay is completely absent and replaced by quite irregular dynamics, which do not seem to reach an equilibrium [$N_\text{eq}$ is set as large as possible \cite{note} in these cases].
These curves hint at the presence of localization effects, which prevent the ``particle'' [in the tight-binding picture] to leave the first site.\\
A histogram of the Gaussian vs.\ exponential data can be viewed in Fig.\ \ref{histoall1}. The mean values read $\bar{\epsilon}_\text{g}=0.15$ in the Gaussian case and $\bar{\epsilon}_\text{e}=0.12$ in the exponential case, which is some orders of magnitude larger than before when shorter wavelengths were excluded. Further, the accumulation of points along the diagonal in the inset suggests that the curves no longer resemble their original form, which is expected judging from \mbox{Fig.\ \ref{allf}}.\\
The histogram for the oscillating cases is displayed in \mbox{Fig.\ \ref{histoall2}}. The mean deviations read $\bar{\epsilon}_\text{gdo}=0.035$ and $\bar{\epsilon}_\text{edo}=0.025$. 
These values are comparable to the Gaussian case in Fig.\ \ref{histo2}. However, since the quantifier $\epsilon$ measures more or less the deviation ``per time step'', the comparison of values $\bar{\epsilon}$ between dynamics that do and do not equilibrate can be void of meaning. In practice, judging from all figures displaying exemplary curves, a value $\epsilon$ of about one percent corresponds to fits that ``look good to the eye''.\\

\noindent
Based on these observations, we specify a property of perturbations that lead to non-generic or pathological dynamics. Namely, a perturbation is  ``untypical'', if it gives rise to Lanczos coefficients that vary non-smoothly, i.e., there is no minimal correlation length within the coefficients. Again, what this means on the level of Hermitian matrices is difficult to assess and beyond the scope of this work. This is the third main result of the paper at hand.

\section{Conclusion}
\label{conc}

\noindent
In this paper, we performed numerical experiments to probe the stability of two classes of relaxation dynamics. The first class consisted of exponentially damped oscillations, which also includes exponential decays. The second class was chosen as a Gaussian counterpart to the first class, i.e., including Gaussian decays and oscillations damped by a Gaussian.\\
The whole strategy was formulated in the framework of the recursion method, in particular, the perturbations were constructed as an alteration of the Lanczos coefficients. Unperturbed coefficients $b_n$ and perturbation $v_n$ were chosen to satisfy certain physically motivated conditions.\\
The first main message of the paper at hand is that
the exponential class of dynamics is relatively stable under the considered perturbations. In contrast, the Gaussian counterpart is found to be quite unstable. These findings confirm and extent upon previous results based on random matrices \cite{knip}, which did not comply with the operator growth hypothesis.\\
We want to emphasize that within this work the main focus should be put on the stability of the former class of relaxation dynamics, as these are ubiquitous in nature, examples are given at the beginning of Sec.\ \ref{numeric}.\\
\noindent
The investigation of the latter class should just be taken as an exemplary comparison. In fact, the choice of a Gaussian counterpart to the first class is quite arbitrary. Any number of relaxation dynamics could have been investigated instead.\\
The second main message of the paper at hand concerns 
the nature of the perturbations themselves. Not only, but also in the context of the works on ``typicality of perturbations'' \mbox{\cite{dabe1,dabe2,dabe3}}, it could be interesting to find  properties of perturbations that lead to non-generic, ``pathological'' dynamics. 
Here, we identified such a criterion, namely that the
perturbation should yield Lanczos coefficients whose minimal correlation length is still above some threshold value.\\
As is evident from Fig.\ \ref{allf},
including short correlations seems to lead to unorthodox dynamics. The displayed curves do not seem to reach an equilibrium, at least on the available time scale. Rather, localization-like effects are introduced, which cause some part of the wave function to remain on the first site. 
On the other hand, sufficiently smooth coefficients alter the relaxation dynamics in a ``controlled yet non-trivial'' manner.\\
Perturbations in various numerical investigations based in random matrices \cite{knip,dabe1} as well as spin lattice models \cite{hev2,jonas} seem to naturally possess the above specified property. Hence, a more systematic investigation on the existence of a minimal correlation length in the coefficients for realistic setups could be a possible prospect for future research.
\newpage

\section*{Acknowledgments}
\label{ack}
\noindent
This work was supported by the Deutsche
Forschungsgemeinschaft (DFG) within the Research Unit
FOR 2692 under Grant No. 397107022 (GE 1657/3-2) and No. 397067869 (STE 2243/3-2).
\newpage

\bibliography{literature}

\begin{thebibliography}{23}
\bibitem [{not()}]{note}%
  \BibitemOpen
  \href@noop {} {\bibinfo  {journal} {In practice, we define the equilibration
  time as follows. We choose a threshold value of $0.005$. If the absolute
  value of $C(t)$ is smaller than the threshold value over the course a certain
  time interval [we choose three time units], the end-time corresponds to the
  upper bound in Eq.\ \eqref{error} and \mbox{Eq.\ \eqref{error2}}. The time
  step is $\delta t=0.01$. If the correlation function does not equilibrate at
  all, $N_\text{eq}$ is set to the maximal time before an echo due to the
  finiteness of the system occurs}\ }\BibitemShut {NoStop}%
\end{thebibliography}

\appendix
\section{Reverse-engineering $b_n$ from $C(t)$}
\label{reveng}
\noindent
In this section, we present the procedure to reverse-engineer the Lanczos coefficients $b_n$ from a given correlation function $C(t)$. The idea is to employ the Lanczos algorithm from Sec.\ \ref{set}, but on the level of spectral functions $\Phi(\omega)$ rather than on the level of observables $|\mco)$. The inner product of operators turns into an inner product of functions, i.e.,
\begin{equation}
(\Phi_1|\Phi_2) = \int \Phi_1^\star(\omega) \Phi_2(\omega)\,\text{d}\omega\,.
\end{equation}
Further, the application of the Liouvillian $\mcl$ to an operator $\mco$ corresponds to a multiplication of the respective function in Fourier space with $-\omega$ [since the commutator with $\mch$ corresponds to a time derivative, which is equivalent to a multiplication with $\ii\omega$ in Fourier space],
\begin{equation}
\mcl|\mco) \rightarrow -\omega \Phi(\omega)\,.
\end{equation}
In practice, we choose a specific correlation function $C(t)$ and calculate its Fourier transform $\Phi(\omega)$. The initial ``seed'' function is set to the (normalized) $\sqrt{\Phi(\omega)}$. Then, the Lanczos algorithm operates as laid out in Sec.\ \ref{set}. In this manner, about $50$ coefficients can be obtained before numerical instabilities become too pronounced.

\section{Link to spectral width}
\label{appb}
\noindent
The eigenvalue equation for the Hamiltonian reads $\mch|E_i\rangle=E_i|E_i\rangle$, where $E_i$ and $|E_i\rangle$ denote eigenvalues and eigenstates, respectively. The corresponding eigenvalue equation for the Liouvillian superoperator is given by $\mcl\mathcal{M}_\beta = \mce_\beta\mathcal{M}_\beta$ with eigenvalues $\mce_{\beta(i,j)}=E_i-E_j$ and eigenoperators $\mathcal{M}_{\beta(i,j)}=|E_i\rangle\langle E_j|$. Without loss of generality, we can set $\text{Tr}[\mch] = \sum_i E_i= 0$. Further, we denote the dimension of the Hilbert space by $d_\mch$, i.e., $d_\mch^{\hspace*{1px}2}=d$.
 Then we have
\begin{align}
\label{spectral}
\sum_n b_n^2 &= \dfrac{1}{2} \text{Tr}[\mcl^2] = \dfrac{1}{2} \sum_{i,j} (E_i-E_j)^2\\
&=\dfrac{1}{2} \sum_{i,j} (E_i^2-2E_iE_j+E_j^2)\nonumber\\\nonumber &=\dfrac{d_\mch}{2} \big{(}\sum_{i} E_i^2 + \sum_j E_j^2  \big{)}\\
 &= d_\mch \text{Tr}[\mch^2] \nonumber
\end{align}
Thus, we see that the quantity in question is indeed linked to the spectral width of the Hamiltonian.

\section{Scaling with perturbation strength}
\label{appc}
\noindent
Consider a Hamiltonian $\mch=\mch_0 + \lambda \mcv$ consisting of an unperturbed part $\mch_0$ and a perturbation $\lambda\mcv$ with $\text{Tr}[\mch_0\mcv]=0$. For the spectral variance of $\mch$ it holds true that
\begin{equation}
\label{ops}
\text{Tr}[\mch^2]=\text{Tr}[\mch_0^2]+\lambda^2 \text{Tr}[\mcv^2]\,.
\end{equation}
On the other hand, with the particular choice of perturbation in Eq.\ \eqref{pert1}, we get  that
\begin{align}
\label{bns}
\sum_n \tilde{b}_n^2 &= \sum_n (b_n+\lambda v_n)^2\\
&= \sum_n (b_n^2+2\lambda b_n v_n + \lambda^2 v_n^2)\nonumber\\ &\approx \sum_n b_n^2 + \lambda^2  \sum_n v_n^2 \nonumber\,.
\end{align}
The term $2\lambda b_nv_n$ is negligible due to the specific choice of $v_n$.
Recalling the relation in Eq.\ \eqref{spectral}, we see from \mbox{Eq.\ \eqref{bns}} that the scaling in Eq.\ \eqref{pert1} reproduces the scaling in Eq.\ \eqref{ops} and is therefore appropriate.

\end{document}